\documentclass[%
      aps,
      prd,
      floatfix,
      preprintnumbers,
      preprint,
      showpacs,
      nofootinbib,
      tightenlines,
      amssymb,
      amsmath
]{revtex4-1}

\usepackage{url}
\usepackage{ifpdf}
\ifpdf
	\usepackage{cmap} 
	\usepackage[pdftex]{color,graphicx}
	\usepackage{epstopdf} 
\else
	\usepackage{color,graphicx}
\fi
\usepackage{bm}
\usepackage[dvipsnames]{xcolor}
\usepackage{hyperref}
\hypersetup{%
colorlinks=true,%
citecolor=Blue,%
filecolor=Black,%
linkcolor=Blue,%
urlcolor=Violet
}
\ifpdf
\usepackage[activate={true,nocompatibility},final,tracking=true,kerning=true,spacing=true,factor=1100,stretch=10,shrink=10]{microtype}
\fi

\newcommand{\eq}[1]{Eq.(\ref{#1})}

\begin{document}
\title{
A Novel Approach to Neutrino-Hydrogen Measurements 
}

\author{H.~Duyang, B.~Guo, S.R.~Mishra and R.~Petti}
\email[]{Roberto.Petti@cern.ch}
\affiliation{Department of Physics and Astronomy, University of South Carolina, Columbia, South Carolina 29208, USA}


\begin{abstract}

The limited statistics of the available (anti)neutrino-hydrogen (H) interactions  has been a longstanding impediment for high-energy neutrino physics. 
We discuss a practical way to achieve accurate (anti)neutrino-hydrogen measurements, addressing the principal limitations of earlier experiments. 
Interactions on hydrogen are extracted by subtracting measurements on thin dedicated graphite (pure C) and polypropylene (CH$_2$) targets 
within a highly segmented low-density detector. A kinematic selection is used to increase the purities to 80-95\% before subtraction. A statistics 
of ${\cal O}(10^6)$ can be realistically achieved in modern neutrino beams for the various $\nu(\bar \nu)$-H event topologies.
The availability of such samples would allow a precise determination of neutrino and antineutrino fluxes, as well as to directly constrain nuclear effects  
from a comparison with corresponding measurements on heavy materials within the same detector. The (anti)neutrino fluxes and the nuclear smearing are 
typically the leading sources of systematic uncertainties in long-baseline oscillation experiments.
(Anti)neutrino-hydrogen interactions also provide an ideal tool for a wide range of precision tests of fundamental interactions.   

\end{abstract}


\maketitle

\section{Introduction}
\label{sec:intro}

Neutrino experiments typically use massive nuclear targets, which are particularly relevant in long-baseline oscillation experiments given the 
reduced flux at the far sites~\cite{Abi:2020evt,Hyper-Kamiokande:2018ofw}. An understanding of the structure and interactions of hadrons within the nuclear 
targets is, therefore, crucial to achieve accurate measurements of neutrino interactions. The existing uncertainties in the modeling of the nuclear effects, which modify the 
neutrino cross-sections, as well as the final state interactions within the nucleus are insufficient for the precisions required by next-generation 
neutrino experiments~\cite{Alvarez-Ruso:2017oui}.

Modern (anti)neutrino beams can deliver intense fluxes allowing the use of high resolution detectors with a relatively modest fiducial mass of a few tons 
to obtain a more accurate reconstruction of (anti)neutrino interactions. However, to exploit the physics potential of such developments for precision measurements 
of fundamental interactions and searches for new physics beyond the Standard Model, neutrino detectors have to address a number of issues. In addition to a high 
experimental resolution and a large acceptance, they must achieve an accurate calibration of the energy scales, as well as a control of the configuration, chemical composition, 
and mass of the neutrino targets comparable to electron scattering experiments~\cite{Petti:2019asx}. 
The fact that the energy of the projectile (anti)neutrino is unknown on an event-by-event basis still represents an intrinsic limitation, 
as neutrino detectors have to infer the (anti)neutrino energy from the reconstructed final state particles emerging from the nucleus, 
which are affected by a substantial nuclear smearing and related systematic uncertainties. The difficulties above are illustrated by 
the many outstanding discrepancies among different experiments, and with existing theoretical models~\cite{Alvarez-Ruso:2017oui}. 
We could therefore argue that the availability of a hydrogen target -- the only hadron target of known energy -- is necessary to go beyond 
the precision level of existing neutrino scattering experiments~\cite{Petti:2022bzt,Petti:2019asx}. 

The available data from $\nu(\bar \nu)$-H interactions were collected by the early bubble chamber experiments 
ANL 12-foot~\cite{Barish:1977qk}, BNL 7-foot~\cite{Fanourakis:1980si}, FNAL E31~\cite{Derrick:1981zb} 
and E45~\cite{Bell:1978ta}, CERN WA21~\cite{WA21:1990dkk} and WA24~\cite{BEBCTSTNeutrino:1983vyc}. 
In spite of the excellent experimental resolution of those measurements, the overall statistics is limited to a total of about 
16k $\nu$-H and 9k $\bar \nu$-H CC interactions. Since then safety requirements related to the underground operation and 
practical considerations have prevented new measurements of high statistics $\nu(\bar \nu)$-H samples. 

In this paper we discuss a novel approach for precision measurements of $\nu(\bar \nu)$-H Charged Current (CC) interactions, which 
is both safe and inexpensive to implement~\cite{Petti:2022bzt,Petti:2019asx,Petti:2023osk}. Interactions on hydrogen are extracted by subtracting measurements 
on dedicated graphite (pure C) targets from those on CH$_2$ plastic targets, integrated within a low-density high resolution detector. 
The concept appears to be a viable and realistic alternative to liquid H$_2$ detectors.

This paper is organized as follows. In Sec.~\ref{sec:targets} we discuss the detection technique and the basic concept of the subtraction 
between CH$_2$ and C targets. In Sec.~\ref{sec:Hsele} we present a detailed kinematic selection of $\nu(\bar \nu)$-H interactions from the CH$_2$ 
plastic target for various event topologies, and in Sec.~\ref{sec:disc} we discuss our results.

\section{Detection technique} 
\label{sec:targets} 

A detector technology designed to offer a control of the configuration, chemical composition, and mass of the neutrino targets similar to electron scattering 
experiments is a Straw Tube Tracker (STT), in which the targets are physically separated from the actual tracking system~\cite{Petti2004}. 
The target mass is distributed within a relatively large volume ($\sim 40$ $m^3$) with the average density similar to that of liquid 
deuterium $\rho \leq 0.17$ $g/cm^3$ and all dimensions comparable to one radiation legth, to achieve an accurate reconstruction of the four-momenta 
of the visible final state particles, as well as of the event kinematics in a plane transverse to the beam direction. 
A large number (70-100) of thin planes -- each typically 1-2\% of radiation length $X_0$ -- of various passive materials with comparable thickness are alternated and 
dispersed throughout active layers -- made of four straw planes -- of negligible mass in order to guarantee the same acceptance 
to final state particles produced in (anti)neutrino interactions. The STT allows to minimize the thickness of individual active layers 
and to approximate the ideal case of a pure target detector, as the targets constitute about 97\% of the total mass~\cite{Petti:2022bzt,Petti:2019asx,Petti:2023osk}. 
The lightness of the tracking straws and the chemical purity of the targets, together with the physical spacing among the individual target planes, 
make the vertex resolution ($\ll 1$ $mm$) less critical in associating (anti)neutrino interactions to the correct target material. 
Each target plane can be removed or replaced with different materials during data taking, providing a flexible target configuration.  

The detector considered here is based upon a central STT inserted in a 0.6 T magnetic field for the measurement of charged-particle momenta, 
and surrounded by a $4\pi$ electromagnetic calorimeter (ECAL)~\cite{Adinolfi:2002zx} for the detection of photons and neutral hadrons including neutrons. 
The base tracking technology is provided by low-mass straws similar to the ones used in many modern experiments for precision physics or the 
search for rare processes~\cite{Sergi:2012twa,Nishiguchi:2017gei,Lee20162530,Anelli:2015pba,Abat:2008zz,PANDA:2013jpu}. 

In this paper we will focus on the ``solid" hydrogen technique, in which $\nu(\bar \nu)$ interactions on free protons are obtained by subtracting measurements 
on dedicated graphite (C) targets from those on polypropylene (CH$_2$) targets~\cite{Petti:2022bzt,Petti:2019asx}. 
The latter are based on one of the plastic materials with the highest hydrogen content (14.4\% by mass) and can be easily manufactured in thin foils of arbitrary size. 
Both C and CH$_2$ targets must have comparable thickness in terms of radiation and nuclear interaction lengths, and are alternated throughout the entire 
tracking volume to ensure that they result in a difference between their detector acceptances for final state particles within $10^{-3}$~\cite{Petti:2023osk}. 
Given the chemical purity achievable for the thin passive targets ($\sim$100\%) and the accuracy in associating the interactions to each target, the  
normalization of the H signal and the C background is based upon the relative abundances in the CH$_2$ compound. 
The technique is conceived to be model-independent, as the data from the graphite targets automatically include all 
types of processes, as well as detector effects, relevant for the selection of interactions on H. Furthermore, it can be safely implemented to obtain 
a relatively large ($\sim 0.7$ ton with 5 tons of CH$_2$) fiducial mass of hydrogen. In the following we will present detailed studies of the corresponding 
event selection with realistic assumptions for the detector smearing and the physics modeling (Sec.~\ref{sec:framework}).

\section{Selection of \boldmath $\nu(\bar \nu)$-H Interactions} 
\label{sec:Hsele} 

The subtraction technique described in Sec.~\ref{sec:targets} can be used to select any inclusive and exclusive process in both CC 
and Neutral Current (NC) $\nu(\bar \nu)$ interactions on free protons~\cite{Petti:2022bzt}. For CC interactions we can improve the signal/background ratio in the selection of 
H interactions by exploiting the event kinematics. Since the H target is at rest, CC events are expected to be perfectly balanced in a plane transverse to the 
beam direction (up to the tiny beam divergence) and the muon and hadron vectors are back-to-back in the same plane. Instead, events 
from nuclear targets are affected by the smearing with the energy-momentum distribution of bound nucleons (Fermi motion and binding), 
the off-shell modifications, meson exchange currents and nuclear shadowing~\cite{Alvarez-Ruso:2017oui,Kulagin:2004ie,Kulagin:2007ju,Kulagin:2014vsa}, 
as well as by final state interactions (FSI)~\cite{Alvarez-Ruso:2017oui}. These nuclear effects result in a significant missing transverse momentum 
and a smearing of the transverse plane kinematics. The use of transverse plane variables and event kinematics to select various Neutral 
Current (NC) and CC (anti)neutrino topologies was pioneered by the NOMAD experiment~\cite{Altegoer:1997gv,Astier:2001yj,Naumov:2004wa,Astier:2000ng}. 
The analysis described in this paper is largely based upon the variables and techniques developed and validated with 
NOMAD data~\cite{Astier:2001yj,Naumov:2004wa,Astier:2000ng}. 

\begin{figure}[tb]
\begin{center} 
\includegraphics[width=1.00\textwidth]{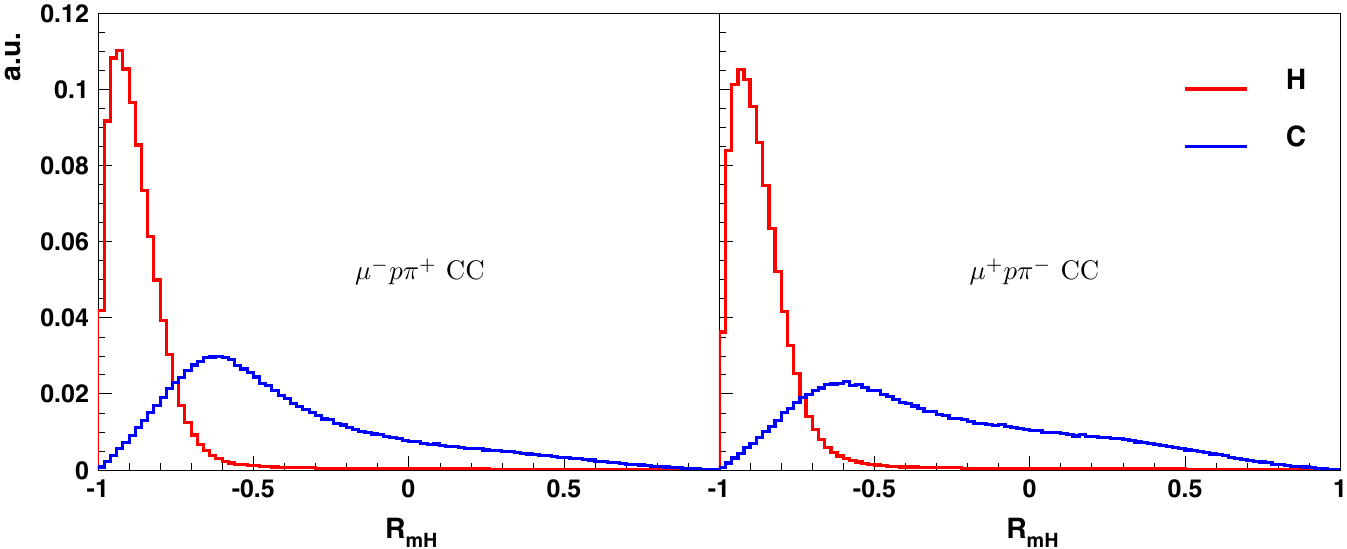}
\end{center} 
\caption{Comparison of the reconstructed $p_T$ asymmetry 
$R_{mH} = (p_T^m - p_T^H)/(p_T^m + p_T^H)$ in 
$\nu_\mu p \to \mu^- p \pi^+$ and $\bar \nu_\mu p \to \mu^+ p \pi^-$ processes 
on H (signal) and in the corresponding background topologies from the C nucleus. 
Results for both neutrino (left panel) and antineutrino (right panel) interactions are shown. All distributions are normalized to unit area. 
} 
\label{fig:Rmh} 
\end{figure} 

\subsection{Analysis framework} 
\label{sec:framework} 

We simulate (anti)neutrino interactions on CH$_2$, H, and C targets with three different event generators: NuWro~\cite{Juszczak:2005zs}, 
GiBUU~\cite{Buss:2011mx}, and GENIE~\cite{Andreopoulos:2009rq}. While we use NuWro as our default generator, we compare our results with both 
GiBUU and GENIE to check the sensitivity of our analysis to the details of the input modeling. These generators are based upon different assumptions and 
modeling of nuclear effects and final state interactions, with GiBUU using a conceptually different approach from heavy ion physics, 
based upon the Boltzmann-Uehling-Uhlenbeck equation for the particle propagation through the nuclear medium. For a detailed review and 
comparison of the generators used we refer to Ref.~\cite{Mosel:2019vhx}. The studies of nuclear FSI by the MINER$\nu$A and T2K 
experiments~\cite{Lu:2018stk,Abe:2018pwo} have found an unphysical excess of $hA$ elastic scattering processes in the FSI simulated by GENIE, which is in 
disagreement with (anti)neutrino data. We follow the corresponding prescriptions by MINER$\nu$A and T2K and discard such elastic $hA$ FSI processes in GENIE. 
We generate inclusive CC interactions including all processes available in the event generators -- quasi-elastic (QE), $\Delta(1232)$ and higher resonances (RES), 
non-resonant processes and deep inelastic scattering (DIS) -- with defaults settings and the input (anti)neutrino spectra expected at the future Long-Baseline 
Neutrino Facility (LBNF)~\cite{Abi:2020evt,Rout:2020cxi,LBNF-flux}. 

We use the GEANT4~\cite{Agostinelli:2002hh} simulation package to evaluate detector effects and apply to all final state particles a parameterized reconstruction 
smearing consistent with the NOMAD data~\cite{Altegoer:1997gv,Anfreville:2001zi}. The detector smearing has also been independently checked 
with the FLUKA~\cite{fluka2} simulation program. The acceptance for individual final state particles ($p,n,\pi^\pm, \pi^0,\mu$) takes into account the detector 
geometry, the event topology, and the material traversed by the particles and is folded into the analysis. We emphasize that the STT detector has been explicitly 
designed to offer the same acceptance for particles produced in both the CH$_2$ and graphite targets, as discussed in Sec.~\ref{sec:targets}. 

For charged tracks the average momentum resolution is about 5\% and the angular resolution about 2 mrad. The reconstruction of protons is typically 
worse due to the shorter track length, with average momentum resolution of 6.5\% for H events and 8.4\% for C events. The difference between H and C events 
is related to the different momentum and angular distributions introduced by nuclear effects. The vertex resolution can vary from 100 $\mu m$ to about 
600 $\mu m$ for multi-track events, depending upon the geometry and the event topology~\cite{Altegoer:1997gv,Anfreville:2001zi}. 
As discussed in Sec.~\ref{sec:targets}, this parameter is not critical in STT, as even for events with a single reconstructed charged track
(Sec.~\ref{sec:numubarqe}) for which the vertex cannot be reconstructed, the uncertainty in associating the event to the correct target material 
is given by the ratio between the thickness of the straw walls ($<20 \mu m$) and the thickess of a single target layer, typically below 0.5\%.
Particle identification is provided by various methods including dE/dx, range, and transition radiation in STT, as well as the energy deposition, timing and 
track segments in the surrounding ECAL and muon identifier. 

We analyze $\nu_\mu$ and $\bar \nu_\mu$ CC interactions originated from the CH$_2$ and graphite targets described in Sec.~\ref{sec:targets}. We determine 
the momentum vectors of charged particles from the track curvature in the B field, while for neutral particles we use either the energy deposited in the 
various sub-detectors or, whenever available, the secondary charged tracks originated by the neutral particles in STT. 
The momentum vector of the total hadron system, $\vec{p}_H$, is obtained from the sum of the momenta of all the reconstructed final state hadrons~\cite{Astier:2001yj}. 

In the following section we discuss a unified approach to the kinematic selection of all the event topologies available in $\nu_\mu$-H and $\bar \nu_\mu$-H CC 
interactions. For more details about the analysis technique used we refer to Ref.~\cite{Astier:2001yj}.

\subsection{Kinematic analysis} 
\label{sec:kine} 

\subsubsection{Selection of $\nu_\mu {\rm H} \to \mu^- p \pi^+$ 
and $\bar \nu_\mu {\rm H} \to \mu^+ p \pi^-$} 
\label{sec:3trk} 

In order to illustrate the potential of the proposed technique, we start from an analysis of the cleanest topologies $\nu_\mu p \to \mu^- p \pi^+$ and 
$\bar \nu_\mu p \to \mu^+ p \pi^-$, mainly originating from resonance production. Detailed GEANT4 simulations indicate that the average proton reconstruction 
efficiency~\footnote{To be reconstructed the proton track emerging from a vertex must have at least four hits in the bending plane to allow a momentum determination. 
We do not make any attempt to reconstruct events failing such a requirementfrom the energy deposition of the available hits.} 
is about 93\% for H events and 67\% in C events, since in this latter case nuclear effects result, on average, in smaller kinetic energies and larger angles for the protons.
As discussed in Sec.~\ref{sec:framework}, the final state particles from these processes on H can be accurately reconstructed in the detector described in 
Sec.~\ref{sec:targets}, thus resulting in an excellent measurement of all the relevant kinematic variables. 

\begin{figure}[tb]
\begin{center} 
\includegraphics[width=1.00\textwidth]{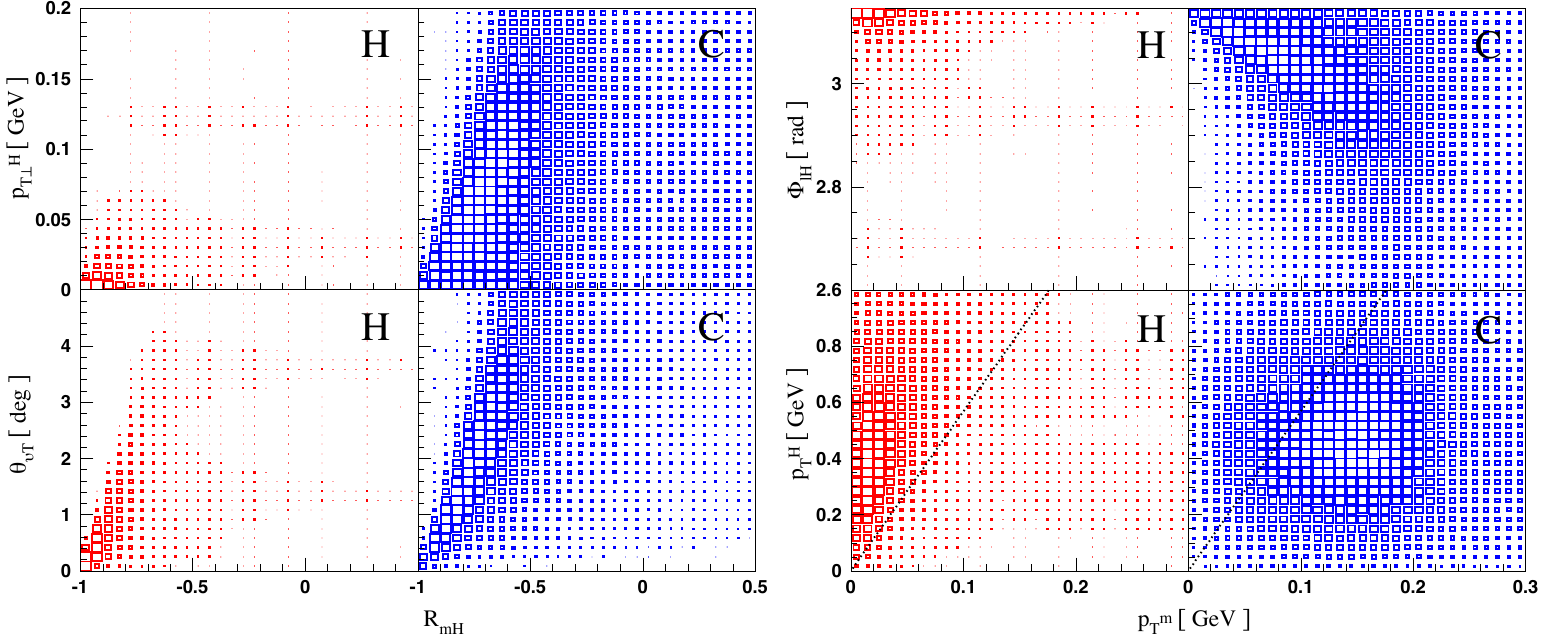}
\end{center} 
\caption{Some of the correlations between kinematic variables used to construct $\ln \lambda^H$ 
and $\ln \lambda_4^H$ for H signal (red color) and C background (blue color) 
exclusive $\mu^- p \pi^+$ CC topologies. The dotted line on the bottom right plots corresponds to a 
cut $R_{mH}<-0.7$ (Fig.~\ref{fig:Rmh}). 
} 
\label{fig:Hsel-Hvar} 
\end{figure} 

The most discriminant kinematic variable to separate H interactions from the ones originated in nuclear targets is found to be 
$R_{mH} \equiv (p_T^m - p_T^H)/(p_T^m + p_T^H)$, the asymmetry between the missing transverse momentum, $p_T^m$, and the transverse momentum 
of the hadron vector, $p_T^H$. For H interactions $p_T^m$ is consistent with zero up to reconstruction effects and hence we expect $R_{mH} \sim -1$. 
Instead, if the interactions occur inside a nuclear target we expect, on average, a substantial $p_T^m$ together with smaller values of $p_T^H$, due to the nuclear smearing.
Furthermore, the missing transverse momentum is mainly generated inside the hadron system and it is expected to be correlated with the latter. 
All these nuclear effects result in much larger values of $R_{mH}$. As shown in Fig.~\ref{fig:Rmh}, this variable can be efficiently used to separate H interactions, 
as well as to probe various aspects of the nuclear smearing. Another useful variable is the magnitude of the component of the hadron transverse momentum 
perpendicular to the transverse momentum of the lepton, $p_{T \perp}^H$. In H interactions the transverse momenta of the lepton and of the hadron system are back-to-back, 
thus resulting in a a sharp peak around zero in $p_{T \perp}^H$. Interactions from nuclear targets have a much broader distribution originating from the 
nuclear smearing. The use of this variable to study H interactions within composite targets was suggested in Ref.~\cite{Lu:2015hea}. Since $p_{T \perp}^H$ is selecting 
topologies in which the transverse momenta of the lepton and the hadron system are back-to-back, the effect of this variable is similar to the use of the angle between 
those transverse vectors, $\Phi_{lH}$. Used as a single variable in the H selection, $p_{T \perp}^H$ has less discriminating power than both $R_{mH}$ and the other 
kinematic variables discussed in this section. However, it provides information complementary to $R_{mH}$, so that the combined effect 
of both these variables improves the overall selection efficiency. 

\begin{figure}[tb]
\begin{center} 
\includegraphics[width=1.00\textwidth]{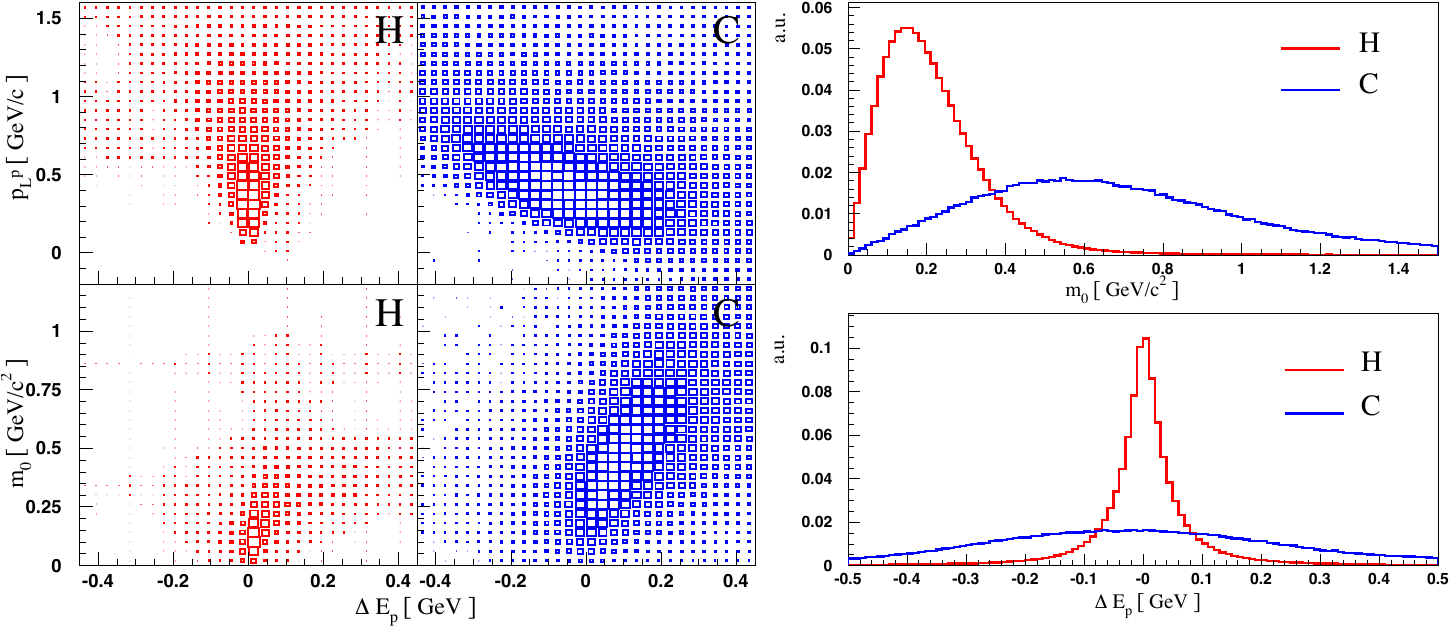}
\end{center} 
\caption{Some of the kinematic variables and correlations used to 
construct $\ln \lambda^H_{\rm IN}$ for H signal (red color) and C background (blue color) 
exclusive $\mu^- p \pi^+$ CC topologies. All histograms in the right plots are normalized to 
unit area.} 
\label{fig:varin} 
\end{figure} 

We can improve the selection of H interactions by using multivariate techniques exploiting the complete event kinematics~\cite{Astier:2001yj,Astier:2000ng,Naumov:2004wa}. 
Assuming the two momentum vectors of the lepton and hadron system, we have in total three transverse and two longitudinal degrees of freedom in the event selection, due to 
the invariance for an arbitrary rotation in the transverse plane. Since we want to separate the same type of CC events with and without nuclear effects, we can further 
assume that the overall reconstructed energy spectra are similar, thus somewhat reducing the rejection power of one of the longitudinal variables compared to the 
transverse ones. As a result, we can define a complete kinematic set as three transverse plus one longitudinal variables. We select the angle between the total visible 
momentum vector and the incident neutrino direction ($z$ axis), $\theta_{\nu T}$, as the variable including longitudinal information. This variable is expected to be 
close to zero in H interactions, up to the tiny beam divergence, while it is much larger in interactions originated from nuclear targets. 

We use likelihood functions incorporating multi-dimensional correlations among kinematic variables. A study of the kinematic selection suggests the following 
function using only three-dimensional correlations: 
\begin{equation} 
{\cal L}^{\rm H} \equiv \left[ [\;R_{mH}, \; p_{T \perp}^H, \; \theta_{\nu T} \;], 
\; p_T^m, \; \Phi_{lH} \; \right]
\label{eq:lnH} 
\end{equation} 
where the square brackets denote correlations (Fig.~\ref{fig:Hsel-Hvar}). A function strictly based upon a complete set of kinematic variables is the four-dimensional: 
\begin{equation} 
{\cal L}^{\rm H}_4=[\; p_T^H, \; p_T^m, \; \Phi_{lH} , \; \theta_{\nu T} \; ]
\label{eq:lnH4} 
\end{equation} 
which incorporates the $R_{mH}$ variable through its underlying correlation $[\; p_T^H, \; p_T^m \;]$ (Fig.~\ref{fig:Hsel-Hvar}). 
Although the use of the ${\cal L}^{\rm H}_4$ function requires a larger statistics to achieve a sensible binning, we used it for our kinematic selection and obtained 
results similar to ${\cal L}^{\rm H}$.
We build the ${\cal L}^{\rm H}$ and ${\cal L}^{\rm H}_4$ probability density functions (pdf) for the two test hypotheses of H interactions (signal) and C interactions (background). 
The individual pdf are properly smoothed and are built with samples independent from the test ones to avoid large statistical biases. As it is common practice, the logarithm 
of the final likelihood ratio between signal and background hypotheses, $\ln \lambda^{\rm H}$ or $\ln \lambda_4^{\rm H}$, is used as discriminant~\cite{Astier:2001yj}. 
We note that in the actual measurements we will directly use interactions from the graphite targets to build the pdf for the C hypothesis, eliminating thus any dependence 
from the nuclear modeling. 

\begin{figure}[tb]
\begin{center} 
\includegraphics[width=1.00\textwidth]{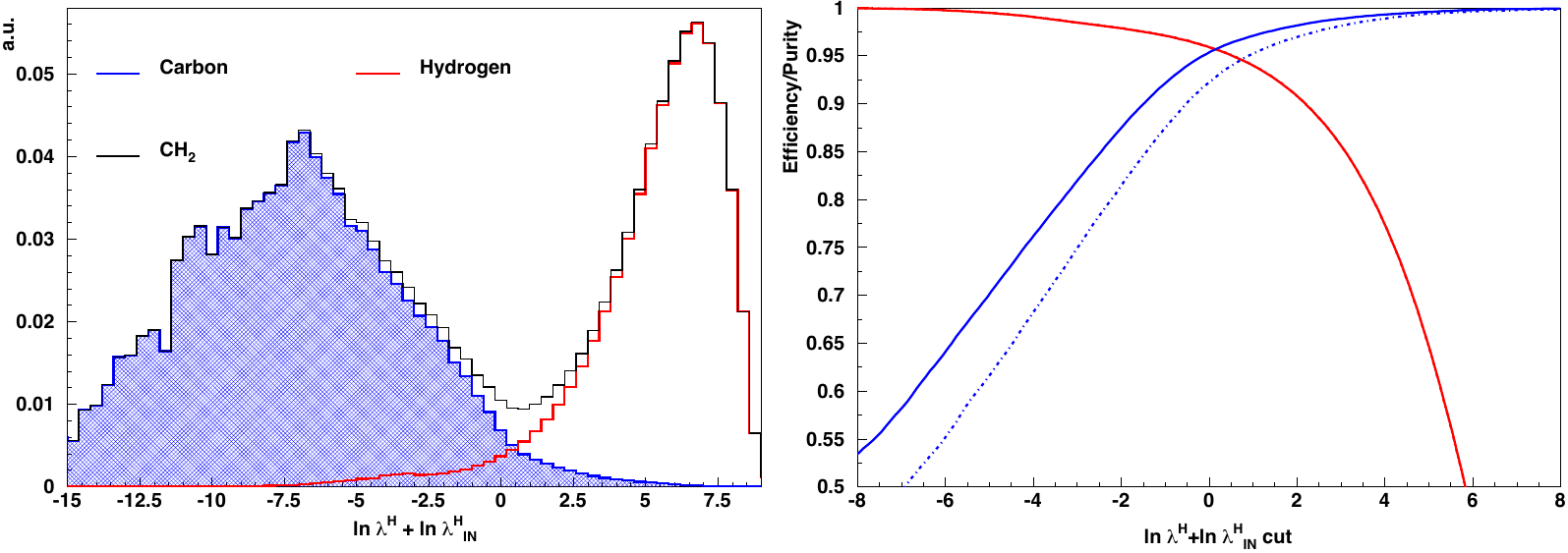}
\end{center} 
\caption{Left plot: Distributions of $\ln \lambda^{\rm H}$+$\ln \lambda^{\rm H}_{\rm IN}$ 
for the H signal, the C background, and the CH$_2$ plastic (sum) for the 
exclusive $\mu^-p\pi^+$ CC topologies. The multiple peaks result from the binning used to build ${\cal L}^{\rm H}$ and ${\cal L}^{\rm H}_{\rm IN}$. 
The C distribution is normalized to unit area while the H distribution is normalized to 
the expected relative abundance in CH$_2$. 
Right plot: Efficiency (red color) and purity (blue color) as a function of the cut on 
$\ln \lambda^{\rm H}$+$\ln \lambda^{\rm H}_{\rm IN}$ for the kinematic 
selection of the exclusive processes $\nu_\mu p \to \mu^-p\pi^+$ 
(solid lines) and $\bar \nu_\mu p \to \mu^+p\pi^-$ (dashed-dotted lines) 
on hydrogen from the CH$_2$ target. The efficiency curves are common to both channels. 
} 
\label{fig:Hlike} 
\end{figure} 

We can also exploit the information related to the individual particles inside the hadron system. For $\mu^\mp p \pi^\pm$ topologies we have a total of three 
additional degrees of freedom, since the total hadron momentum vector is constrained by the global event kinematics in ${\cal L}^{\rm H}$ and ${\cal L}_4^{\rm H}$. 
We select one such variable as the difference $\Delta E_p$ between the energy of the neutrino calculated from the $\mu$ and $\pi$ momenta by applying 
energy-momentum conservation and the one reconstructed from the sum of the momentum vector of all three particles $\mu^\mp p \pi^\pm$: 
\begin{equation} 
\Delta E_p = \frac{m_\mu^2-m_{\pi^\pm}^2+2M_p\left( E_\mu + E_{\pi^\pm} \right) - 
2p_\mu \cdot p_{\pi^\pm}} 
{2\left( M_p - E_\mu - E_{\pi^\pm} + \mid \vec{p}_\mu \mid \cos \theta_\mu + 
\mid \vec{p}_{\pi^\pm} \mid \cos \theta_{\pi^\pm} \right)} - 
\mid \vec{p}_\mu + \vec{p}_{\pi^{\pm}} + \vec{p}_p \mid
\label{eq:RES1} 
\end{equation} 
where $M_p, M_n, m_\mu, m_{\pi^\pm}$ are the masses of the proton, neutron, muon, and pion, respectively, $p_{\mu (\pi^\pm)}, \vec{p}_{\mu (\pi^\pm)}, E_{\mu (\pi^\pm)}$ 
and $\theta_{\mu (\pi^\pm)}$ are the four-momentum, momentum vector, energy and angle of the outgoing muon (pion), and $\vec{p}_p$ is the proton momentum vector. 
The variable $\Delta E_p$ is close to zero up to reconstruction effects in hydrogen, while it is significantly larger in carbon events, due to the nuclear smearing. 
Another useful variable is the invariant mass of the reconstructed neutrino, calculated as 
$m_0^2 = ( p_\mu + p_{\pi^\pm} + p_p - p_N )^2$ 
where $p_p$ and $p_N$ are the four-momenta of the outgoing proton and of the target proton assumed at rest, respectively. 
We use the following likelihood function using information from the internal $p \pi$ structure: 
\begin{equation} 
{\cal L}^{\rm H}_{\rm IN}=[\; \Delta E_p, \; p_L^p, \; m_0 \;]
\label{eq:lnHin} 
\end{equation} 
where $p_L^p$ is the longitudinal component of the proton momentum vector along the beam direction. Figure~\ref{fig:varin} shows the main variables and 
correlations included in ${\cal L}^{\rm H}_{\rm IN}$. Since ${\cal L}^{\rm H}_{\rm IN}$ is essentially independent from ${\cal L}^{\rm H}$ and ${\cal L}^{\rm H}_{4}$ we  
multiply the corresponding density functions and use the sum $\ln \lambda^{\rm H}$+$\ln \lambda^{\rm H}_{\rm IN}$ or $\ln \lambda^{\rm H}_4$+$\ln \lambda^{\rm H}_{\rm IN}$ 
as the final discriminant for our analysis. 

\begin{table}[tb] 
\begin{center} 
\begin{tabular}{|l|c|c|c|c|c|c|c|c|c|c|} \hline
    &     \multicolumn{4}{|c|}{ $\nu_\mu$-H CC} 
    &     \multicolumn{6}{|c|}{$\bar \nu_\mu$-H CC}  \\ 
Process  &  $\mu^- p \pi^+$  &  $\mu^- p \pi^+ X$ & $\mu^- n \pi^+ \pi^+ X$ &  Inclusive  &  
 $\mu^+ p \pi^-$ & $\mu^+ n \pi^0$ & $\mu^+ n$ & $\mu^+ p \pi^- X$ & $\mu^+ n \pi \pi X$ &  Inclusive \\ 
\hline\hline
Eff. $\varepsilon$ & 96\% & 89\% & 75\% & 93\% & 94\% & 84\% & 75\%& 85\% & 82\% & 80\% \\  
Purity & 95\% & 93\% &  70\%& 93\%  & 95\% & 84\% & 80\% & 94\% & 84\% & 84\% \\ 
\hline
\end{tabular}
\caption{Efficiency $\varepsilon$ and purity for the kinematic selection of H interactions from the CH$_2$ plastic target using the likelihood ratio 
$\ln \lambda^{\rm H}$+$\ln \lambda^{\rm H}_{\rm IN}$ 
or $\ln \lambda^{\rm H}_4$+$\ln \lambda^{\rm H}_{\rm IN}$. For the $\mu^+ n$ QE 
topologies $\ln \lambda^{\rm H}_{\rm QE}$ is used instead. 
The cuts applied for each channel are chosen to maximize the sensitivity defined 
as $S/\sqrt{S+B}$, where $S$ is the H signal and $B$ the C background. The CC inclusive samples 
are obtained from the combination of the corresponding exclusive channels. 
} 
\label{tab:effpur} 
\end{center} 
\end{table} 

The distributions of $\ln \lambda^{\rm H}$+$\ln \lambda^{\rm H}_{\rm IN}$ for the H signal and the C background in  $\mu^- p \pi^+$ topologies are shown 
in Fig.~\ref{fig:Hlike} (left plot). The corresponding purity and efficiency achievable as a function of the $\ln \lambda^{\rm H}$+$\ln \lambda^{\rm H}_{\rm IN}$ cut 
are given in the right plot of Fig.~\ref{fig:Hlike}, for both the $\nu_\mu p \to \mu^- p \pi^+$ and $\bar \nu_\mu p \to \mu^+ p \pi^-$ samples. 
Both the efficiency and the purity appear relatively uniform as a function of the neutrino energy. Table~\ref{tab:effpur} summarizes the results obtained by applying 
the cut on $\ln \lambda^{\rm H}$+$\ln \lambda^{\rm H}_{\rm IN}$ maximizing the sensitivity $S/\sqrt{S+B}$, where $S$ is the number of events for H signal and $B$ 
for the C background. The fact that the maximum sensitivity corresponds to regions with high purity for the selected H signal indicates that the kinematic selection is optimal. 
An advantage of this approach is that the likelihood function allows to evaluate, on an event-by-event basis, the probability that a given (anti)neutrino interaction 
originated from either the hydrogen or the carbon nucleus. Furthermore, it provides a better control of the selection procedure with respect to a simple cut-based analysis, 
by allowing an easier variation of the efficiency/purity and by offering relatively clean control samples. 

\begin{figure}[tb]
\begin{center} 
\includegraphics[width=1.00\textwidth]{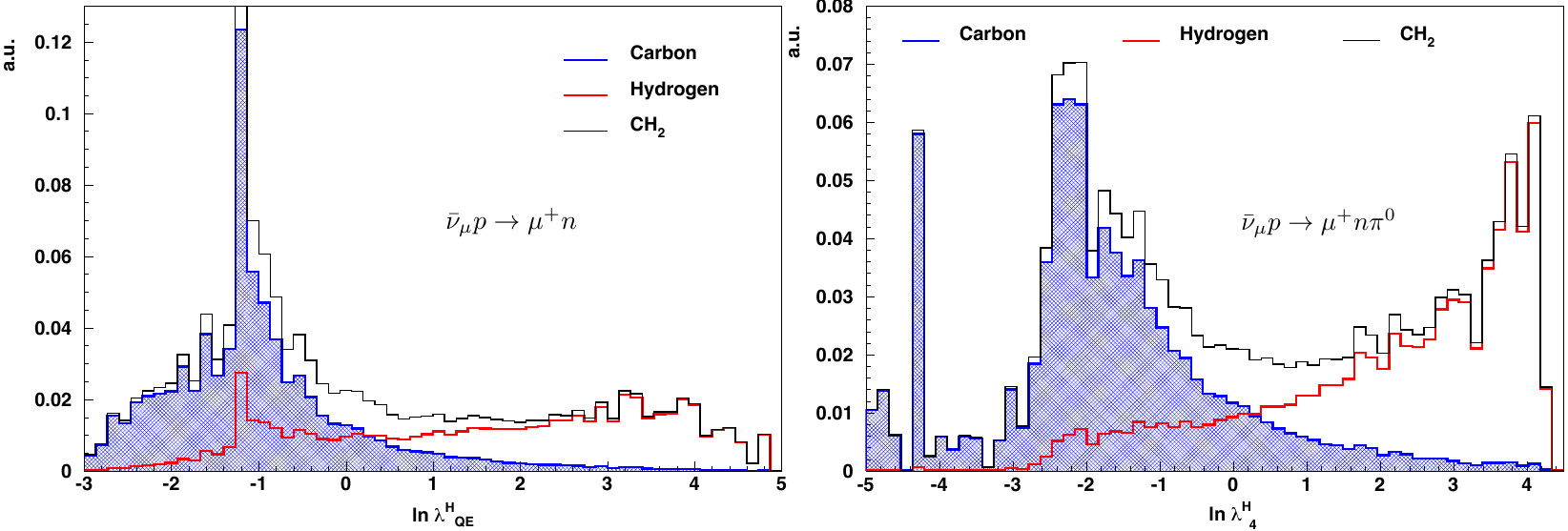}
\end{center} 
\caption{Left plot: Distributions of $\ln \lambda^{\rm H}_{\rm QE}$ for 
the H signal, the C background, and the CH$_2$ plastic (sum) for 
the selection of $\bar \nu_\mu p \to \mu^+ n$ QE events.
Right plot: Distributions of $\ln \lambda^{\rm H}_4$ 
for the selection of $\bar \nu_\mu p \to \mu^+ n \pi^0$ events. 
The multiple peaks are the effect of the binning used to build the likelihood functions. 
The C distributions are normalized to unit area while the H distributions are normalized to 
the expected relative abundance in CH$_2$. See the text for details. 
} 
\label{fig:Hsel-likeHqepi0} 
\end{figure}

\subsubsection{Selection of $\bar \nu_\mu {\rm H} \to \mu^+ n$} 
\label{sec:numubarqe} 

An important exclusive process available in $\bar \nu_\mu$ CC interactions is the quasi-elastic $\bar \nu_\mu p \to \mu^+ n$ on hydrogen. The reconstruction of 
this topology is more complex because of the presence of a neutron in the final state and a single charged track. We performed detailed GEANT4 simulations of 
the detector to study the reconstruction efficiency of the neutrons. The results were independently checked using FLUKA simulations. 
About 25\% of the neutrons produced in $\bar \nu_\mu p \to \mu^+ n$ QE events on H interact inside the STT and are detected from the corresponding hits in the straws. 
An additional fraction of about 55\% of the neutrons are detected in the ECAL~\cite{Anelli:2009zza} surrounding the tracker, thus allowing a combined detection 
efficiency of 80.5\% for H events. Conversely, the combined detection efficiency for neutrons originated from the C background events is only about 64.8\%, since 
they are typically affected by nuclear effects, resulting, on average, in a smaller kinetic energy. 
For events with a single charged track the resolution on the position of the primary vertex is worse than for multi-track events, and is essentially defined by the 
thickness of a single CH$_2$ (or C) target plane by noting the absence of straw tube hits preceding the presumed target. 
However, events can still be efficiently associated to the correct target material, due to the lightness of the tracking straws and the purity of the target itself. 
The corresponding uncertainty is given by the ratio between the thickness of the straw walls ($< 20$ $\mu m$) and the thickness of a single CH$_2$ target, 
resulting in an efficiency $>99\%$ (Sec.~\ref{sec:framework}). From the position of the first track hit and that of the neutron interaction within the detector we can 
reconstruct the neutron line of flight. Assuming that the target proton is at rest, we calculate the energy of the incoming antineutrino as: 
\begin{equation} 
E_\nu = \frac{M_n^2-m_\mu^2+2M_pE_\mu -M_p^2}{2\left(  M_p- E_\mu +\mid \vec{p}_\mu \mid \cos \theta_\mu\right)}
\label{eq:QE} 
\end{equation} 
with the same notations as in \eq{eq:RES1}. The energy of the neutron is $E_n=E_\nu+M_p-E_\mu$ and the momentum vector of the neutron is obtained from the 
measured neutron direction and the calculated energy $E_n$~\footnote{It is possible to calculate the complete momentum vector of the neutron from energy-momentum conservation. However, the direct use of transverse plane kinematics would void the kinematic selection against the C background.}. 
We note that \eq{eq:QE} is correct only for interactions on hydrogen and not for the ones on carbon, due to nuclear effects. We use a realistic smearing on the measured 
direction of the neutron obtained from a detailed FLUKA simulation of the detector and including the non-gaussian tails, mainly related to elastic scattering of the 
neutrons before interacting. We can then reconstruct the complete event kinematics for $\bar \nu_\mu p \to \mu^+ n$ interactions on hydrogen and apply the same 
kinematic selection described in Sec.~\ref{sec:kine}. The resolution achievable on many of the kinematic variables is dominated by the reconstruction of the neutron 
direction rather than that of the $\mu^+$ momentum vector, given the excellent momentum and angular resolutions of the STT. 

An additional background source to consider for the selection of the process $\bar \nu_\mu p \to \mu^+ n$ on H is given by the uncorrelated neutrons 
originated from (anti)neutrino interactions occurring in the large amount of materials surrounding the STT, including the ECAL, the magnet elements, 
and the external rocks. We estimate this background by randomly overlaying a genuine $\bar \nu_\mu p \to \mu^+ n$ event on H with a neutron extracted 
from either a different CH$_2$ event or from an event in the surrounding materials within the same beam spill. We then reconstruct the neutron momentum using 
the measured direction of the random neutron and the calculated energy $E_n$, as described above. We require a small time difference between the interaction 
point of the random neutron and the primary vertex, $\Delta t \equiv t_{\rm n} - t_{\rm vtx}$. We also require that the ``measured" time-of-flight $\Delta t$ is consistent 
with the expected one using the calculated neutron energy, with small values of the variable $\Delta t^\prime \equiv \Delta t - d/(\beta_n c)$. 
We further reject the random neutrons which are closely correlated -- in time and space -- to a detected activity from an external (anti)neutrino interaction detected 
either in STT or in the surrounding ECAL. The combined effect of these timing and topological cuts rejects most of the random neutrons by retaining 90.4\% of H events. 

\begin{figure}[tb]
\begin{center} 
\includegraphics[width=0.50\textwidth]{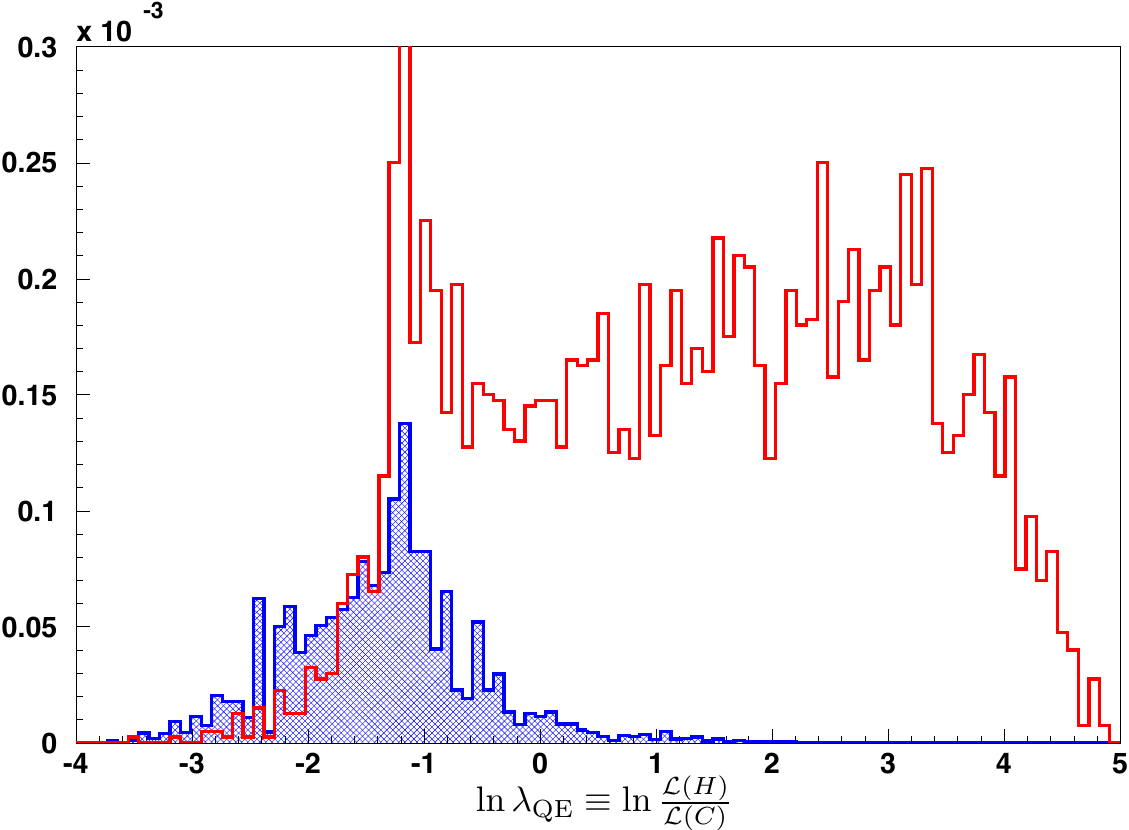}
\end{center} 
\caption{Distribution of $\ln \lambda_{\rm QE}^{\rm H}$ for the random neutron background and for the 
$\bar \nu_\mu p \to \mu^+ n$ events on H. The likelihood functions calculated from the C and H hypotheses 
are used, as in Fig.~\ref{fig:Hsel-likeHqepi0}. The multiple peaks are the effect of the binning used 
to build the likelihood functions. Timing and topological cuts are applied as described in the text. 
} 
\label{fig:nrandom} 
\end{figure} 

We have less kinematic information available for the selection of $\bar \nu_\mu p \to \mu^+ n$ QE on H with respect to other channels. 
We use the optimal function ${\cal L}^{\rm H}_4$ from \eq{eq:lnH4} as the basis of our kinematic analysis. The likelihood function ${\cal L}^{\rm H}_{\rm IN}$, 
as defined in \eq{eq:lnHin}, is not strictly applicable, since the hadron system is composed of a single neutron without internal structure. 
However, we can still exploit one additional longitudinal degree of freedom with respect to \eq{eq:lnHin}.
In analogy of \eq{eq:RES1}, we calculate $\Delta E_n = E_\nu \; - \mid \vec{p}_\mu + \vec{p}_n \mid $, where $E_\nu$ is given by \eq{eq:QE} and 
$\vec{p}_n$ is the neutron momentum calculated as described above. We use the following likelihood function for the selection of $\bar \nu_\mu p \to \mu^+ n$ QE on H: 
\begin{equation} 
{\cal L}^{\rm H}_{\rm QE}= \left[ [\; p_T^H, \; p_T^m, \; \Phi_{lH} , \; \theta_{\nu T} \; ], 
 \; \Delta E_n, \;  p_L^n \;  \right]
\label{eq:QE2} 
\end{equation} 
The variable $p_L^n$ represents the longitudinal component of the neutron momentum vector along the beam direction. We note that, although the two-dimensional correlation 
 $ [\; \Delta E_n, \;  p_L^n \;]$ appears similar to the corresponding one in \eq{eq:lnHin}, it has a different role in the QE channel, as a part of the global event kinematics. 
We therefore include it into a single likelihood function, since it is expected to be correlated with the other kinematic variables. 
Figure~\ref{fig:Hsel-likeHqepi0} shows the distributions of $\ln \lambda_{\rm QE}^{\rm H}$ for the H signal and the C background. Table~\ref{tab:effpur} summarizes 
the efficiency and purity for a cut on $\ln \lambda_{\rm QE}^{\rm H}$ maximizing the sensitivity of the analysis. Since the random neutron background is uncorrelated 
with the primary vertex from the $\bar \nu_\mu p \to \mu^+ n$ interaction on H, the corresponding kinematic distributions are inconsistent with those expected from 
a genuine H event. Figure~\ref{fig:nrandom} shows the distribution of $\ln \lambda_{\rm QE}^{\rm H}$ for the random neutrons passing the timing and topological 
cuts described above. We do not re-calculate the likelihood functions using the random neutron events and we keep the same $\ln \lambda_{\rm QE}^{\rm H}$  
cuts as in the H/C separation, resulting in no additional efficiency loss for the H signal. We note that the efficient kinematic selection could allow 
looser timing and topological cuts against random neutrons.

\subsubsection{Selection of $\bar \nu_\mu {\rm H}  \to \mu^+ n \pi^0$} 
\label{sec:numubarpi0} 

Another exclusive process available only in $\bar \nu_\mu$ CC interactions is $\bar \nu_\mu p \to \mu^+ n \pi^0$ on hydrogen, mainly originated from 
resonance production. The kinematic selection of this channel is similar to the one discussed in Sec.~\ref{sec:3trk} for the complementary process 
$\bar \nu_\mu p \to \mu^+ p \pi^-$, with the exception of a few specific features related to the reconstruction of the neutron and the $\pi^0$. 
The relative fractions of these two channels are about comparable in $\bar \nu_\mu$ CC interactions on hydrogen. Detailed GEANT4 simulations indicate that the 
average neutron reconstruction efficiency is about 85\% for H events and 76.5\% in C using both the STT and the surrounding ECAL (Sec.~\ref{sec:numubarqe}). 

For the reconstruction of the $\pi^0$ we distinguish the case in which $\gamma$s are converted into $e^+e^-$ pairs within the STT volume and the one in 
which $\gamma$s are detected only in the surrounding ECAL. In the former case we use the momentum and anglular smearing for $e^\pm$ tracks in STT, 
while in the latter case we apply the corresponding smearing obtained from a detailed ECAL simulation. On average, about 50\% of all $\pi^0$ have at least one 
converted $\gamma$ within the STT volume, providing a more accurate reconstruction of the $\pi^0$ direction. We calculate the energy of the neutrons 
interacting in the detector as: 
\begin{equation} 
E_n = \frac{M_n^2-m_\mu^2-m_{\pi^0}^2+2M_p\left( E_\mu + E_{\pi^0} \right) - 2p_\mu \cdot p_{\pi^0}-M_p^2} 
{2\left( M_p - E_\mu - E_{\pi^0} + \mid \vec{p}_\mu \mid \cos \theta_\mu + 
\mid \vec{p}_{\pi^0} \mid \cos \theta_{\pi^0} \right)} + M_p - E_\mu - E_{\pi^0}
\label{eq:RES2} 
\end{equation} 
where $p_{\pi^0}, \vec{p}_{\pi^0}, E_{\pi^0)}$ and $\theta_{\pi^0}$ are the four-momentum, momentum vector, energy and angle of the outgoing $\pi^0$, respectively.  
The first term on the r.h.s. of \eq{eq:RES2} represents the neutrino energy calculated from the muon and $\pi^0$ momenta using energy-momentum conservation. 
We note that \eq{eq:RES2} is correct only for interactions on hydrogen since it assumes a target proton at rest. We reconstruct the neutron momentum vector $\vec{p}_n$ 
following the same procedure described in Sec.~\ref{sec:numubarqe} for the QE process, combining the measured (smeared) direction of the neutron 
with the energy $E_n$ calculated from \eq{eq:RES2}. 

We use the likelihood function ${\cal L}^{\rm H}_4$ from \eq{eq:lnH4} to describe the global event kinematics. In analogy to the $\bar \nu_\mu p \to \mu^+ p \pi^-$ case, 
it is possible to exploit the additional information related to the individual particles within the hadron system with the function 
${\cal L}^{\rm H}_{\rm IN}= [\; \Delta E_{\pi^0}, \; p_L^{\pi^0}, \; m_0 \;]$, where $\Delta E_{\pi^0} = E_\nu - \mid \vec{p}_\mu + \vec{p}_{\pi^0} + \vec{p}_n \mid$,  
$E_\nu$ is the neutrino energy calculated from the muon and neutron momenta using energy-momentum conservation, and $p_L^{\pi^0}$ is the longitudinal 
component of the momentum vector of the $\pi^0$. The use of $\Delta E_{\pi^0}$ is preferable with respect to the equivalent quantity for the neutron, $\Delta E_n$, 
since this latter is partially biased by the calculation of the neutron energy from \eq{eq:RES2}. Since the average reconstruction smearing for the neutron and $\pi^0$ 
is larger than for most charged particles, the $\ln \lambda^{\rm H}_{\rm IN}$ improves only marginally the selection of $\bar \nu_\mu p \to \mu^+ n \pi^0$ on H and 
can be dropped for this channel. Figure~\ref{fig:Hsel-likeHqepi0} shows the distributions of $\ln \lambda_4^{\rm H}$ for the H signal and the C background. 
Table~\ref{tab:effpur} summarizes the efficiency and purity for a cut on $\ln \lambda_4^{\rm H}$ maximizing the sensitivity of the analysis. 
Similar results are obtained with $\ln \lambda^{\rm H}$.

\subsubsection{Selection of $\nu_\mu {\rm H} \to \mu^- p \pi^+ X$ 
and $\bar \nu_\mu {\rm H} \to \mu^+ p \pi^- X$} 
\label{sec:disp} 

In this section and in the following one we consider the collective selection of all the inelastic topologies produced in $\nu_\mu p$ and $\bar \nu_\mu p$ CC interactions 
on H and different from the $\mu^- p \pi^+$,  $\mu^+ p \pi^-$, and $\mu^+ n \pi^0$ topologies discussed in Sec.~\ref{sec:3trk} and Sec.~\ref{sec:numubarpi0}. 
These samples are dominated by the Deep Inelastic Scattering (DIS): about 63\% of the $\mu^- \pi^+ X$ events and 7\% of the $\mu^- p \pi^+$ 
sample have $W>1.8$ GeV with the default LBNF beam spectra~\cite{LBNF-flux}. 

\begin{figure}[tb]
\begin{center} 
\includegraphics[width=1.00\textwidth]{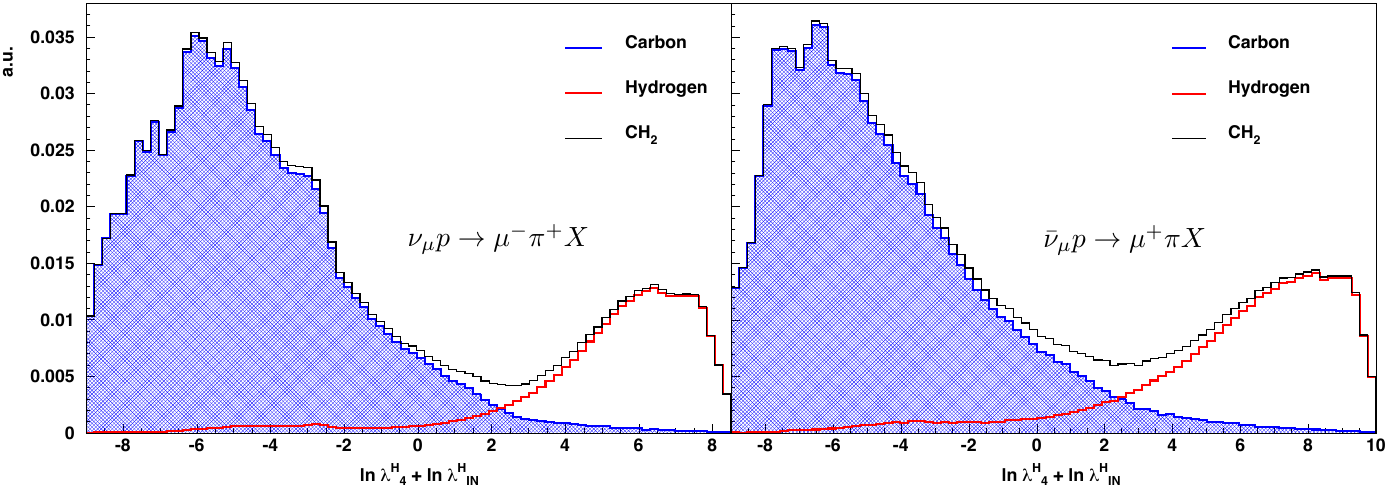}
\end{center} 
\caption{Distributions of $\ln \lambda^{\rm H}_4$+$\ln \lambda^{\rm H}_{\rm IN}$ 
for the H signal, the C background, and the CH$_2$ plastic (sum) for 
$\nu_\mu p \to \mu^- \pi^+ X$ (left plot) and $\bar \nu_\mu p \to \mu^+ \pi X$ (right plot), 
including the sum of the subsamples with a proton and a neutron. 
The multiple peaks are the effect of the binning used to build 
${\cal L}_4^{\rm H}$ and ${\cal L}^{\rm H}_{\rm IN}$. 
The C distributions are normalized to unit area while the H distributions are normalized to 
the expected relative abundance in CH$_2$. See the text for details. 
} 
\label{fig:Hsel-likeHdis} 
\end{figure} 

The inelastic $\nu_\mu p \to \mu^- \pi^+ X$ and $\bar \nu_\mu p \to \mu^+ \pi X$ samples on H are characterized, in general, by a higher multiplicity and a worse 
reconstruction of some of the events, compared to the $\mu^- p \pi^+$,  $\mu^+ p \pi^-$, and $\mu^+ n \pi^0$ topologies. The larger detector smearing directly 
affects the reconstruction of the kinematic variables, thus somewhat reducing their discriminating power. However, these effects are even larger for the DIS interactions 
originated in nuclear targets, primarily because of the final state interactions. We can therefore still achieve an adequate separation of the interactions on hydrogen 
from the carbon background, which largely dominates the statistics of the $\mu^- \pi^+ X$ and $\mu^+ \pi X$ samples from the CH$_2$ target.

Since we want to select interactions on protons, we require that the total charge measured at the primary vertex is $C=+1$. This cut rejects a large fraction of the 
interactions on neutrons in background events and has an efficiency of about 23\%(33\%) for $\nu_\mu (\bar \nu_\mu)$ CC interactions on C. 
We also require the presence of at least one $\pi^+$ in $\nu_\mu p$ and one $\pi^-$  or $\bar \nu_\mu p$ events to select inelastic interactions. 

In $\nu_\mu p$ and $\bar \nu_\mu p$ CC interactions on H we expect a single nucleon -- either a proton or a neutron -- in the final state, while for interactions on 
C nuclear effects including final state interactions can result in higher multiplicities. We perform a separate analysis of the two subsamples with a proton and a neutron. 
In this section we discuss the selection of the subsamples with a reconstructed proton. Detailed GEANT4 simulations indicate that the average proton reconstruction 
efficiency is about 96\% for H events and 74\% in C events. 

The kinematic selection of $\nu_\mu p \to \mu^- p \pi^+ X$ and $\bar \nu_\mu p \to \mu^+ p \pi^- X$ on H can exploit additional degrees of freedom with respect 
to the $\mu^- p \pi^+$,  $\mu^+ p \pi^-$, and $\mu^+ n \pi^0$ topologies, due to the higher number of particles in the hadron system. We use the likelihood 
function ${\cal L}^{\rm H}_4$ from \eq{eq:lnH4} to describe the global event kinematics. We then tag the single particle within the hadron system 
potentially affected by the largest nuclear effects by maximizing the magnitude $\mid \Delta E_{h_i} \mid$, defined as: 
\begin{equation} 
\Delta E_{h_i} = 
\frac{m_{h_i}^2-m_\mu^2-(W^{\prime})^2+2M_p\left( E_\mu + E_{H^\prime} \right) - 2p_\mu \cdot p_{H^\prime}-M_p^2} 
{2\left( M_p - E_\mu - E_{H^\prime} + \mid \vec{p}_\mu \mid \cos \theta_\mu + 
\mid \vec{p}_{H^\prime} \mid \cos \theta_{H^\prime} \right)} 
- \mid \vec{p}_\mu + \sum_k  \vec{p}_{h_k}  \mid
\label{eq:DIS} 
\end{equation} 
which is calculated for each hadron particle $h_i$ of mass $m_{h_i}$ and momentum vector $\vec{p}_{h_i}$. The first term on the r.h.s. of \eq{eq:DIS} is similar to 
the equivalent one in \eq{eq:RES2}, with the $\pi^0$ replaced by a reduced hadron system $H^\prime$ excluding the single particle $h_i$ being considered. The quantity 
$(W^\prime)^2=E_{H^\prime}^2-\mid \vec{p}_{H^\prime} \mid^2$ is the invariant mass of the reduced hadron system, and $E_{H^\prime}=\sum_{k\neq i} E_{h_k}$ and 
$\vec{p}_{H^\prime} = \sum_{k \neq i} \vec{p}_{h_k}$ are the corresponding energy and momentum vector. We select the hadron particle maximizing the magnitude 
$\mid \Delta E_{h_i} \mid$ and use the corresponding value $\Delta E_{h_i}^{\rm max}$ (with sign) and the longitudinal momentum of such a particle, $p_L^{h_i}$, 
as input for the likelihood function~\footnote{In principle the same approach can be used for $\mu^- p \pi^+$, $\mu^+ p \pi^-$, and $\mu^+ n \pi^0$ topologies as well.} 
based upon information internal to the hadron system: 
\begin{equation} 
{\cal L}^{\rm H}_{\rm IN}=[\; \Delta E_{h_i}^{\rm max}, \; p_L^{h_i}, \; m_0, \; ] 
\label{eq:DIS2} 
\end{equation} 
where we use similar notations as in \eq{eq:lnHin}. Figure~\ref{fig:Hsel-likeHdis} shows the distributions of $\ln \lambda_4^{\rm H}$+$\ln \lambda_{\rm IN}^{\rm H}$ 
for the H signal and the C background in $\nu_\mu p \to \mu^- \pi^+ X$ and $\bar \nu_\mu p \to \mu^+ \pi^- X$ topologies. Table~\ref{tab:effpur} summarizes the
efficiency and purity in the selection of both $\nu_\mu p \to \mu^- p \pi^+ X$ and $\bar \nu_\mu p \to \mu^+ p \pi^- X$ processes on H with a cut on 
$\ln \lambda_4^{\rm H}$+$\ln \lambda_{\rm IN}^{\rm H}$ maximizing the sensitivity of the analysis. 
Similar results are obtained with $\ln \lambda^{\rm H}$+$\ln \lambda_{\rm IN}^{\rm H}$.

\subsubsection{Selection of $\nu_\mu {\rm H} \to \mu^- n \pi^+ \pi^+ X$ 
and $\bar \nu_\mu {\rm H} \to \mu^+ n \pi \pi X$} 
\label{sec:disn} 

In this section we discuss the selection of the subsamples of the inelastic $\nu_\mu p$ and $\bar \nu_\mu p$ CC interactions on H with a detected neutron in 
the final state and different from the $\mu^+ n \pi^0$ topologies (Sec.~\ref{sec:numubarpi0}). These samples are complementary with respect to the similar ones 
with a reconstructed proton in the final state described in Sec.~\ref{sec:disp}. Detailed GEANT4 simulations indicate that the average neutron reconstruction
efficiency is about 87.4\% for H events and 80.1\% in C events using both the STT and the surrounding ECAL (Sec.~\ref{sec:numubarqe}). For $\nu_\mu p$ CC on H 
the presence of a neutron in the final state is tagged by requiring that no proton and two $\pi^+$ are present. These criteria correctly tag the neutron in 99.8\% of the H 
events with a neutron, including events in which the neutron is not detected. Events with a neutron in $\bar \nu_\mu p$ CC on H are tagged with an efficiency of 
88\% by requiring that no proton and a number of pions ($\pi^-$+$\pi^+$+$\pi^0$) $\geq 2$ with total charge equal to zero are present in the final state. 
The analysis follows closely the one described in Sec.~\ref{sec:disp} for the subsamples with a proton, the main difference being the treatment of the neutron. 
We calculate the neutron energy using energy-momentum conservation and \eq{eq:RES2} with the $\pi^0$ replaced by a reduced hadron system $H^\prime$ 
excluding the neutron: 
\begin{equation} 
E_n = \frac{M_n^2-m_\mu^2-(W^{\prime})^2+2M_p\left( E_\mu + E_{H^\prime} \right) - 2p_\mu \cdot p_{H^\prime}-M_p^2} 
{2\left( M_p - E_\mu - E_{H^\prime} + \mid \vec{p}_\mu \mid \cos \theta_\mu + 
\mid \vec{p}_{H^\prime} \mid \cos \theta_{H^\prime} \right)} + M_p - E_\mu - E_{H^\prime}
\label{eq:DIS3} 
\end{equation} 
where we use the same notations as in \eq{eq:DIS} with $h_i\equiv n$. We reconstruct the neutron momentum vector $\vec{p}_n$ following the same procedure 
described in Sec.~\ref{sec:numubarqe} for the QE process, combining the measured (smeared) direction of the neutron with the energy $E_n$ calculated from \eq{eq:DIS3}. 
For events with more than one neutron detected the calculation above is not applicable and we ignore the neutrons. 

We use ${\cal L}^{\rm H}_4$ from \eq{eq:lnH4} to describe the global event kinematics and ${\cal L}^{\rm H}_{\rm IN}$ from \eq{eq:DIS2} for the 
information related to the individual particles inside the hadron system. Since the angular smearing for the detected neutrons (Sec.~\ref{sec:numubarqe}) 
is typically larger than for other particles, we use the track with the largest angle with respect to the beam direction to calculate $ \Delta E_{h_i}^{\rm max}$ 
in \eq{eq:DIS2}, rather than explicitly maximizing $ \mid \Delta E_{h_i} \mid$. 
Figure~\ref{fig:Hsel-likeHdis} shows the distributions of $\ln \lambda_4^{\rm H}$+$\ln \lambda_{\rm IN}^{\rm H}$ for the H signal and the C background 
in $\nu_\mu p \to \mu^- \pi^+ X$ and $\bar \nu_\mu p \to \mu^+ \pi X$ topologies. Table~\ref{tab:effpur} summarizes the efficiency and purity in the selection of 
both $\nu_\mu p \to \mu^- n \pi^+ \pi^+ X$ and $\bar \nu_\mu p \to \mu^+ n \pi \pi X$ processes on H with a cut on $\ln \lambda_4^{\rm H}$+$\ln \lambda_{\rm IN}^{\rm H}$ 
maximizing the sensitivity of the analysis. Similar results are obtained with $\ln \lambda^{\rm H}$+$\ln \lambda_{\rm IN}^{\rm H}$.

\subsubsection{Selection of $\nu_\mu {\rm H}$ and $\bar \nu_\mu {\rm H}$ CC inclusive} 
\label{sec:inclusive} 

In the previous sections we optimized the selection of the various exclusive topologies available in $\nu_\mu p$ and $\bar \nu_\mu p$ CC interactions on H 
by maximizing independently the corresponding sensitivities. The results summarized in Tab.~\ref{tab:effpur} are characterized by varying efficiencies and purities 
across different channels. For measurements requiring the inclusive CC samples we can combine the individual exclusive topologies with their 
corresponding relative fractions in $\nu_\mu p$ and $\bar \nu_\mu p$ CC interactions on H. The average efficiency and purity of the resulting inclusive 
CC samples on H are listed in Tab.~\ref{tab:effpur}.

\subsection{Achievable statistics} 
\label{sec:stat} 

\begin{figure}[tb]
\begin{center} 
\includegraphics[width=0.70\textwidth]{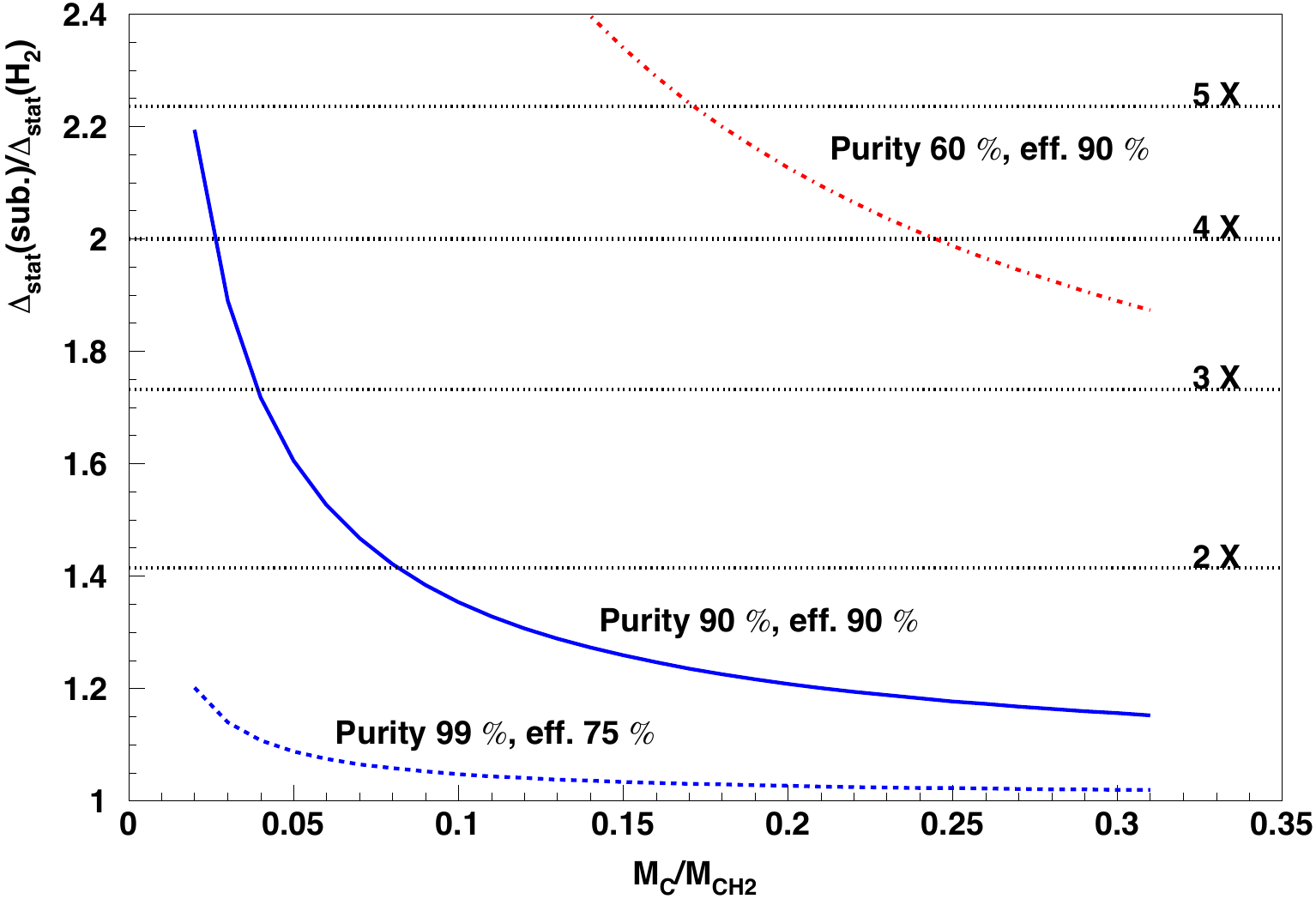}
\end{center} 
\caption{Ratio between the relative statistical uncertainty $\Delta_{\rm stat}$ on the H sample obtained after the C background subtraction and the corresponding ideal 
one from a pure H$_2$ sample of the same size, as a function of the ratio between the fiducial masses of the graphite and CH$_2$ targets, $M_C/M_{CH_2}$. 
The curves for different values of the H purities and efficiencies are shown to illustrate the impact of the kinematic selection. See text for details. 
} 
\label{fig:Hsel-Hstat} 
\end{figure} 

In the following we will assume an overall fiducial mass of 5 tons for the CH$_2$ targets, corresponding to a ``solid" hydrogen mass of about 700 kg. 
This value is realistically achievable with the detector technology discussed in Sec.~\ref{sec:targets} and a relatively compact tracking volume around 40 m$^3$,  
depending upon the specific configuration of the main STT parameters. The distributions of the generic kinematic variables $\vec x \equiv (x_1, x_2, ....., x_n)$ in 
$\nu(\bar \nu)$-H CC interactions are obtained as: 
\begin{equation} 
N_{\rm H}(\vec x) \equiv N_{\rm CH_2} (\vec x) - \frac{M_{\rm C/CH_2}}{M_{\rm C}} N_{\rm C} (\vec x)
\label{eq:Hevt} 
\end{equation} 
where $N_{\rm CH_2}$ and $N_{\rm C}$ are the numbers of events selected from the polypropylene and graphite targets, respectively. The interactions from 
this latter are normalized by the ratio between the total fiducial masses of C within the graphite and CH$_2$ targets, $M_{\rm C/CH_2}/M_{\rm C}$. 
The subtraction in \eq{eq:Hevt} is performed after all the selection cuts, including the kinematic analysis described in Sec.~\ref{sec:Hsele} and resulting in the purities 
and efficiencies summarized in Tab.~\ref{tab:effpur}. Practical considerations require the graphite targets to be smaller than the actual mass of C inside the CH$_2$ plastic, 
thus resulting in a statistical penalty associated with the subtraction procedure. Figure~\ref{fig:Hsel-Hstat} illustrates how the total statistical uncertainty on $N_H$ from 
\eq{eq:Hevt} compares to the ideal one expected from a pure H$_2$ sample equivalent to the statistics of H interactions within CH$_2$. 
For a given efficiency, the purity of the H samples achievable by the kinematic selection can drastically reduce the overall statistical uncertainty from the subtraction technique. 
Our analysis suggests that a fiducial mass for the graphite targets around 600 kg corresponding to $M_C/M_{CH_2}\sim$ 0.12 provides a reasonable compromise with 
a statistical penalty around 30\%. We note that this statistical penalty can be further reduced by analytically smoothing the measured distributions from the graphite target 
and/or by using a tighter kinematic selection, as illustrated in Fig.~\ref{fig:Hsel-Hstat}. 

\begin{table}[tb]
\begin{center} 
\begin{tabular}{|l|r|r|r|r|r|} \hline
CC process  &  ~~CH$_2$  target~~ &   ~~H target~~ &  ~~CH$_2$ selected~~  &  ~~C bkgnd~~  & ~~H selected~~\\ \hline\hline
$\nu_\mu p \to \mu^- p \pi^+$   &  5,615,000   &   2,453,000   & 2,305,000  & 115,000  & 2,190,000 \\ 
$\nu_\mu p \to \mu^- p \pi^+ X$~~~   &  11,444,000  & 955,000 & 877,000 &  61,000 & 816,000 \\
$\nu_\mu p \to \mu^- n \pi^+ \pi^+ X$~~~   &  3,533,000  & 183,000 & 158,000 & 48,000 & 110,000  \\
\hline
$\nu_\mu$ CC inclusive~~  &   34,900,000   &  3,591,000    &  3,340,000   &  224,000  & 3,116,000 \\ 
\hline\hline
$\bar \nu_\mu p \to \mu^+ n$   &   4,450,000    &  1,688,000    &  1,274,000  &  255,000 & 1,019,000 \\
$\bar \nu_\mu p \to \mu^+ p \pi^-$   &   827,000   &  372,000    &  342,000  &  17,000 & 325,000 \\ 
$\bar \nu_\mu p \to \mu^+ n \pi^0$   &  791,000 & 366,000 &  295,000 & 48,000 & 247,000 \\
$\bar \nu_\mu p \to \mu^+ p \pi^- X$   & 2,270,000  &  176,000 & 153,000 & 9,000  & 144,000 \\
$\bar \nu_\mu p \to \mu^+ n \pi \pi X$   & 2,324,000  & 280,000 & 220,000 & 35,000 & 185,000 \\
\hline 
$\bar \nu_\mu$ CC inclusive~~ &   13,000,000   &  2,882,000    & 2,284,000   &  364,000 & 1,920,000   \\ 
\hline
\end{tabular}
\caption{Number of events expected in the selection of all the various processes on H with the default low-energy (anti)neutrino beams available at the  
LBNF~\cite{LBNF-flux}, assuming a fixed exposure of $5.5\times 10^{21}$ POT, a detector located at 574 m from the source and a fiducial target mass of 5 tons of CH$_2$. 
The first two columns (CH$_2$ and H targets) refer to the initial statistics, while the last three include all selection cuts described in Sec.~\ref{sec:Hsele} and Tab.~\ref{tab:effpur}). 
For the CH$_2$ and C targets the numbers refer to the given final state topologies originated from either $p$ or $n$ interactions. The fifth column shows the total residual C 
background to be subtracted from the corresponding CH$_2$ selected samples. We use a ratio $M_C/M_{C/CH_2} =0.12$ to measure the C backgrounds from the graphite targets. 
See text for details. 
} 
\label{tab:events} 
\end{center} 
\end{table} 

We consider the beam spectra expected in the LBNF project~\cite{LBNF-flux} and assume a fixed exposure of $5.5\times 10^{21}$ protons on target (POT) for both the 
low-energy neutrino and antineutrino beams, achievable in about two years with the default beam power of 1.2 MW.  
Table~\ref{tab:events} summarizes the total number of events expected for the various topologies and targets. An interesting option available at LBNF is a 
high-energy beam optimized to detect the $\nu_\tau$ appearance from neutrino oscillations in the far detector, which would result in an increase by a factor 2.4 of the 
$\nu(\bar \nu)$-H rates, combined with a much harder spectrum. It is conceivable to have a dedicated two year run with such a high energy beam after the completion of 
the nominal data taking. By that time the planned upgrades of the beam intensity to a nominal power of 2.4 MW would further roughly double the available POT.

\section{Discussion} 
\label{sec:disc} 

\subsection{Systematic uncertainties} 
\label{sec:syst}

The kinematic analysis described in Sec.~\ref{sec:kine} allows the identification of all the various $\nu(\bar \nu)$-H CC topologies within the CH$_2$ target in STT 
with little residual backgrounds (5-20\%) from interactions on the carbon nucleus. The kinematic selection can reduce the statistical uncertainty from the background 
subtraction procedure (Sec.~\ref{sec:stat}) and the impact of systematic uncertainties on the modeling of nuclear effects in carbon~\cite{Alvarez-Ruso:2017oui}. 
These latter are further suppressed by a model-independent background subtraction using the data obtained from the dedicated graphite target~\cite{Petti:2022bzt,Petti:2019asx},  
which also provides a pure background sample to build the likelihood functions used in the selection. 
The detector technology discussed in Sec.~\ref{sec:targets} allows the integration of a large number CH$_2$ and C targets, which are configured as thin (1-2\% $X_0$) 
passive layers with the same equivalent thickness in terms of radiation and nuclear interaction lengths, and are alternated throughout the detector volume.
Detailed detector simulations indicate that in this way the acceptance difference between targets can be kept within $10^{-3}$ for all particles~\cite{Petti:2023osk}. 
The data from the graphite target automatically include all types of interactions, as well as reconstruction effects, relevant for our analysis. 

\begin{table}[tb]
\begin{center} 
\begin{tabular}{|l|c|c|c|c|c|c|c|} \hline
    &   &  \multicolumn{2}{|c|}{ NuWro} 
    &     \multicolumn{2}{|c|}{GiBUU}  &  \multicolumn{2}{|c|}{GENIE}  \\ 
Process  & Selection & ~~Efficiency~~  &  ~~Purity~~  &  ~~Efficiency~~  &  ~~Purity~~  &  ~~Efficiency~~  &  ~~Purity~~ \\ \hline\hline
$\nu_\mu p \to \mu^- p \pi^+$   &  $\ln \lambda^{\rm H}$+$\ln \lambda^{\rm H}_{\rm IN}$ &  96\%  &    95\%  & 96\%   &  85\%   &  96\%  &  96\%  \\ 
$\bar \nu_\mu p \to \mu^+ p \pi^-$   & $\ln \lambda^{\rm H}$+$\ln \lambda^{\rm H}_{\rm IN}$ &  94\%  &    95\%  &  94\%  &  87\%  &  94\%  & 98\% \\
$\bar \nu_\mu p \to \mu^+ p n$ & $\ln \lambda^{\rm H}_{\rm QE}$ &    75\%  &  80\%  &   75\%  &  88\%   &  75\%  &  94\%  \\ 
\hline
\end{tabular}
\caption{Comparison of the efficiency and purity for the kinematic selection of H interactions from the CH$_2$ target (Sec.~\ref{sec:kine}) with the 
NuWro~\cite{Juszczak:2005zs}, GiBUU~\cite{Buss:2011mx}, and GENIE~\cite{Andreopoulos:2009rq} event generators.
The same cuts on the likelihood ratios as in Tab.~\ref{tab:effpur} are used. In all cases the likelihood functions are built from the default NuWro generator and 
are not recalculated when different generators are tested. See text for details. 
} \label{tab:effpurMC} 
\end{center} 
\end{table} 

The kinematic selection of $\nu_\mu$-H and $\bar \nu_\mu$-H CC interactions described in Sec.~\ref{sec:kine} relies upon the fact that the target proton is at rest 
and on energy-momentum conservation, rather than on the specific kinematics of the interactions with the free nucleon. Uncertainties on these latter, including form 
factors and structure functions, affect only marginally the resulting efficiencies through the corresponding kinematic dependence for individual exclusive topologies. 
Furthermore, uncertainties related to the structure of the free nucleon would be common to both signal and backgrounds, largely canceling in a selection based 
upon the differences introduced by nuclear effects. We note that the relevant proton form factors and structure functions can be directly determined in a 
model-independent way from the measured $Q^2$ and $x$ distributions, using the procedure described in details in Ref.~\cite{Duyang:2019prb}. 

Although the ``solid" hydrogen technique is conceived to be model-independent, in the absence of actual data from STT the efficiencies and purities listed in 
Tab.~\ref{tab:effpur} are sensitive to the details of the interaction modeling implemented in the simulations. In order to estimate the impact of 
such effects, we repeat the event selection described in Sec.~\ref{sec:kine} with three event generators: NuWro~\cite{Juszczak:2005zs}, GiBUU~\cite{Buss:2011mx}, 
and GENIE~\cite{Andreopoulos:2009rq}. These generators use rather different assumptions for the modeling of (anti)neutrino-nucleus interactions 
-- including both initial and final state nuclear effects --as outlined in Ref.~\cite{Mosel:2019vhx}. 
To this end, we do not recalculate the likelihood functions but rather use the ones obtained from the default NuWro generator throughout. 
Given the differences among generators~\cite{Mosel:2019vhx,Duyang:2019prb}, this assumption can help to understand the impact of potential 
discrepancies between data and simulations on the H selection. As shown in Tab.~\ref{tab:effpurMC}, our kinematic selection of H interactions from the  CH$_2$ 
targets is relatively stable across the three event generators. We emphasize that this test is only meant to estimate a possible outer envelope for the numbers in 
Tab.~\ref{tab:effpur}, since the actual technique is entirely data-driven and all backgrounds and efficiencies will be directly determined from the measured interactions. 
Furthermore, we will also use data from the graphite targets to build the likelihood functions independently of simulations. 

Reconstruction effects on the four-momenta of the final state particles can in principle degrade the kinematic selection. For this reason in our studies we used a realistic  
detector smearing and checked its consistency with GEANT4 and FLUKA simulations (Sec.~\ref{sec:framework}). Furthermore, we validated the effects of 
the detector acceptance, smearing, and track reconstruction with NOMAD data~\cite{Altegoer:1997gv}, although the NOMAD 
detector had worse acceptance and granularity than the STT. We note that similar kinematic selections were successfully demonstrated by 
NOMAD~\footnote{The higher neutrino energy in NOMAD implies higher backgrounds and more difficult kinematic selections for low multiplicity processes like 
RES and QE compared to our case study. Furthermore, the lowest usable energy in NOMAD was about 5 GeV, which is reasonably close to the spectra expected at LBNF.} 
in more severe background conditions (rejections up to $10^5$) in various published analyses~\cite{Astier:2001yj,Naumov:2004wa,Astier:2000ng}, as well as in single track 
measurements of $\bar \nu_\mu$ QE and inverse muon decay~\cite{Lyubushkin:2008pe}.

The momentum scale of charged particles can be calibrated with the mass peak of the large samples of reconstructed $K^0 \to \pi^+\pi^-$ decays~\cite{Duyang:2019prb}. 
In our study we assume the same energy scale uncertainty of 0.2\% achieved by the NOMAD experiment using this technique~\cite{Wu:2007ab}. 
We note that the STT at LBNF would provide 25 times higher granularity than NOMAD and about 40 times higher $K^0$ statistics~\cite{Duyang:2019prb}. 
Similarly, the proton identification and reconstruction efficiency can be accurately calibrated with the large samples of $\Lambda \to p \pi^-$ decays available~\cite{Duyang:2019prb}. 
Both $\Lambda$ and $K^0$ decays can be used to constrain the systematic uncertainty on the reconstruction of the track angles. 

The use of a likelihood function in the kinematic analysis (Sec.~\ref{sec:kine}) provides a simple way to vary the purity and efficiency of the selected samples (Fig.~\ref{fig:Hlike}) 
to validate the background subtraction and the selection efficiencies through appropriate control samples~\cite{Astier:2001yj}.

\subsection{Physics Measurements} 
\label{sec:apps} 

The availability of high-statistics samples of $\nu(\bar \nu)$-H CC interactions would be extremely relevant for neutrino scattering physics, as well as for 
long-baseline oscillation experiments. In this section we briefly outline some of the physics measurements~\cite{Petti:2022bzt,Petti:2019asx,Petti2018}. 

The limited knowledge of the (anti)neutrino flux has always been a major limitation for accelerator-based neutrino experiments. The exclusive 
$\nu_\mu {\rm H} \to \mu^- p \pi^+$ and $\bar \nu_\mu {\rm H} \to \mu^+ n$ processes with small energy transfer $\nu$ offer an excellent  tool to measure the relative 
(anti)neutrino flux as a function of energy with little hadronic uncertainties. The use of the samples described in Sec.~\ref{sec:stat} allows a determination of the $\nu_\mu$ 
and $\bar \nu_\mu$ relative flux to a precision better than 1\%~\cite{Duyang:2019prb} in conventional wide-band beams, which is not achievable with other known techniques 
using nuclear targets. The $\bar \nu_\mu {\rm H} \to \mu^+ n$ interactions at small momentum transfer $Q$ also provide an accurate measurement of the absolute 
$\bar \nu_\mu$ flux, since the corresponding cross-section in the limit $Q\to 0$ is known to high accuracy from neutron $\beta$ decay~\cite{Duyang:2019prb,Petti:2023abz}. 

A comparison of $\nu(\bar \nu)$-H CC interactions with the corresponding ones from nuclear targets within the same detector provides a direct measurement 
of nuclear effects~\cite{Alvarez-Ruso:2017oui,Kulagin:2004ie,Kulagin:2007ju,Kulagin:2014vsa}, which typically introduce a substantial smearing of the observed interactions. 
This study can be performed with both inclusive CC events and with various exclusive topologies. Constraining the nuclear smearing from initial and final state interactions 
is required to reduce the systematic uncertainties in the unfolding of data collected from nuclear targets. To this end, the combined use of $\nu$-H and $\bar \nu$-H CC 
interactions provides a control sample free from nuclear effects to calibrate the neutrino energy scale in CC interactions~\cite{Petti:2022bzt}. 

The unique combination of nuclear and ``solid" hydrogen targets would enable a broad program of precision measurements and searches for new 
physics~\cite{Petti:2019asx,Petti:2022bzt} complementary with the ones planned in the collider~\cite{AbdulKhalek:2021gbh},  
fixed target~\cite{Dudek:2012vr}, and nuclear physics communities.  An example is given by the Adler sum rule~\cite{Adler:1964yx}, which is based upon current algebra 
and was tested only by BEBC~\cite{Allasia:1985hw} with a few thousand events. Similarly, by exploiting the isospin symmetry $F_2^{\nu n} = F_2^{\bar \nu p}$, we can obtain 
a direct determination of the free neutron structure functions, as well as a measurement of the large $x$ behavior of the $d/u$ quark ratio~\cite{Alekhin:2017fpf,Alekhin:2022tip}. 
These measurements can also be used for precision tests of the isospin (charge) symmetry~\cite{Petti:2022bzt} and would help to elucidate the flavor structure 
of the nucleon~\cite{Alekhin:2018dbs}. Furthermore, using a combination of both isoscalar and non-isoscalar nuclear targets can provide valuable insights on the physics 
mechanisms responsible of the nuclear modifications of the nucleon properties~\cite{Kulagin:2004ie,Kulagin:2014vsa,Alekhin:2022uwc}. 

\section*{Data Availability Statement} 

No Data associated in the manuscript. 

\begin{acknowledgments}

The authors express their gratitude to the DUNE experiment for the use of some detector simulation tools. We thank L. Camilleri, R. Ent, X. Lu, and X. Qian for fruitful discussions. 
We thank F. Ferraro, L. Di Noto, P. Sala, and M. Torti for providing the reconstruction smearing of $n$ and $\pi^0$ with the FLUKA simulation package.

\end{acknowledgments}

\bibliography{main1}

\end{document}